\documentstyle[aps,multicol,psfig]{revtex}

\def\be{\begin{equation}}
\def\ee{\end{equation}}
\def\bea{\begin{eqnarray}}
\def\eea{\end{eqnarray}}
\def\bean{\begin{eqnarray*}}
\def\eean{\end{eqnarray*}}
\def\half{\frac{1}{2}}
\def\vecnul{{\bf 0}}
\def\vecp{{\bf\rm p}}
\def\veck{{\bf\rm k}}
\def\vecx{{\bf\rm x}}

\def\ep{\epsilon}
\def\lm{\lambda}

\def\om{\omega}
\def\Om{\Omega}
\def\bra{\langle}
\def\ket{\rangle}

\begin{document}

\title{The Classical Approximation for Real-Time Scalar Field Theory at 
Finite Temperature\footnote{Talk presented at the 5th International 
Workshop on Thermal Field Theories and Their Applications, Regensburg, 
Germany, August 10-14, 1998.}}

\author{Gert Aarts}

\address{
Institute for Theoretical Physics, Utrecht University\\
Princetonplein 5, 3508 TA Utrecht, the Netherlands\\
{\rm aarts@phys.uu.nl}}

\date{September 1, 1998}

\maketitle

\begin{abstract}
The use of classical thermal field to approximate real-time quantum 
thermal field theory is discussed. For a $\lambda\phi^4$ theory, it is 
shown that the classical Rayleigh-Jeans divergence can be canceled with 
the appropriate counterterms, and a comparison is 
made between the classical and quantum perturbative expansion. It is 
explained why Hard Thermal Loops  prevent the same method to work for 
gauge theories.
\end{abstract}
  
\begin{multicols}{2}
\section{Introduction}
Real-time dynamics of quantum fields plays an 
important role in the early universe (baryogenesis, inflation) and in heavy 
ion collisions. Some quantities cannot be determined within perturbation 
theory, even after possible resummations. An example is the rate of 
sphaleron transitions at high temperature in (extensions of) the Standard 
Model, relevant for baryogenesis.
A non-perturbative real-time calculation on a lattice, using Monte Carlo 
methods, is complicated, because of the complex Boltzmann weight.

A relatively simple way to do the dynamics is to use the classical 
equations of motion. Depending on the choice of initial conditions, 
this describes an 
equilibrium or a non-equilibrium situation. In this talk I discuss 
classical $\lambda \phi^4$ theory, in thermal equilibrium.
This talk is based on work done in collaboration with Jan Smit 
\cite{AaSm} (see also \cite{JaNa}).
More recent work is presented at this Workshop at the poster session 
\cite{poster}.

\section{Effective field theories and coarse-graining}
It often happens that a Quantum Field Theory (QFT) is too complicated to be 
solved completely. If there are several scales present in the model, a 
useful approach is to construct an effective field theory for the 
degrees of freedom that are important at the scale under consideration. The 
other 'less important' degrees are integrated out, one way or the other. 
This very general idea has been applied successfully in many different 
physical situations, under names as coarse-graining, Wilson renormalization 
group, dimensional reduction, and so on.

These ideas have also been applied 
when considering the real-time dynamics in a QFT at finite temperature. 
This typically leads to a transport or kinetic theory desciption.
As an explicit example, consider real-time $\lambda \phi^4$ theory at 
finite temperature \cite{GrMu96}. The low-momentum or soft  modes are 
considered to be the 
degrees of freedom that are of interest, and the high-momentum or hard 
modes as those that can be integrated out. In general, the resulting 
semi-classical dynamics for the soft modes is very complicated. They
are subject to noise and dissipation, due to the coarse-grained 
interaction with hard modes, and the effective equations of motion are 
non-local in space and time.
Especially this non-locality makes it difficult to use the resulting 
equations directly for a numerical treatment.

Therefore I discuss here another possibility to approximate the 
dynamics, originally proposed in \cite{GrRu88}, 
which is related to dimensional reduction (DR) and 3-d effective field 
theories for static quantities: classical thermal field theory.

\section{Dynamics and classical thermal field theory}
As already mentioned in the Introduction, a relatively simple way to deal 
with 
the dynamics, is to use the classical equations of motions. One simply solves 
the equations, with given initial conditions, and then averages over the 
initial conditions with the Boltzmann weight $\exp -\beta H$ as distribution 
function. If the hamiltonian in the Boltzmann weight is the same as the one 
that 
determines the equations of motion, this leads to an equilibrium description.
For e.g. the classical 2-point function, this procedure means the 
following ($x = (\vecx, t)$) 
\bea
\label{eqS}
 S(x-x') &=& \bra \phi(x)\phi(x')\ket_{\rm cl} \\
\nonumber
&=& Z_{\rm cl}^{-1} \int {\cal D}\pi{\cal D}\phi\,e^{-\beta H(\pi, \phi)}\,
\phi(x)\phi(x'),\eea
with the classical partition function 
\[ Z_{\rm cl} = \int {\cal D}\pi{\cal D}\phi\,e^{-\beta H(\pi, \phi)}, 
\]
$\beta=1/T$, and the hamiltonian and potential
\bean
H &=& \int d^3x\,\half \pi^2 + V(\phi),\\
V &=& \int d^3x\left( \half (\nabla \phi)^2 + \half m^2\phi^2 + 
\frac{\lm}{4!} \phi^4\right).\eean
In (\ref{eqS}), $\phi(x)$ is the solution of the classical equations of 
motion $\dot{\phi}(x) = \{\phi(x), H\}$, $\dot{\pi}(x) = \{\pi(x), H\}$,
with the initial conditions $\phi(\vecx, t_0) = \phi(\vecx)$, $\pi(\vecx, 
t_0) = \pi(\vecx)$. The integration over phase space is over the initial 
conditions at  $t=t_0$, weighted with the Boltzmann weight.

It is necessary to make two remarks at this point. First of all, 
classical thermal field theory contains the well-known Rayleigh-Jeans 
divergence. Hence, everything is formulated with a (lattice) cutoff, to 
regularize this divergence. We will show that this cutoff can be removed 
in the end, if the parameters  in the classical theory are chosen in the 
correct way.
And secondly, when restricting to 
time-independent correlation functions, the canonical momenta can be 
integrated out, and the resulting partition function has precisely the form 
of that of a 3-d, superrenormalizable field theory, 
\[ Z_{\footnotesize\mbox{3-d}} = \int {\cal D}\phi\,e^{-\beta V(\phi)}, 
\]
which has been studied in great detail in DR \cite{DR}. I will come back 
to DR results at the appropriate places.

In the remainder of this talk, I will discuss what role the cutoff and the 
Rayleigh-Jeans divergence play and what the relation with the dynamics in 
the quantum theory is. This will be done in perturbation theory.

\section{Perturbation theory in quantum and classical thermal field theory}
Perturbation theory in the classical theory is obtained by 
writing the field as $\phi(x) =\phi_0(x) +\lambda\phi_1(x)+\ldots$. 
Solving the equations of motion, order by order in $\lambda$, gives
\bean
\phi_0(x) &=& \int \frac{d^3k}{(2\pi)^3}\,e^{i\veck\cdot\vecx}\,[
\phi(\veck)\cos \omega_\veck(t-t_0) \\
&&\;\;\;\;\;\;\;\;+
\frac{\pi(\veck)}{\omega_\veck}\sin \omega_\veck(t-t_0)],\\
\phi_1(x) &=&-\lm\int d^4x'\, 
G^R_0(x-x')\frac{1}{3!}\phi_0^3(x'),\;\;\;\; \mbox{etc},
\eean
with $\omega_\veck^2 = \veck^2+m^2$. Here we introduced the retarded 
Green function
\be
\label{eqGR}
 G^R_{\rm cl}(x-x') =  -\theta(t-t')\bra 
\{\phi(x),\phi(x')\}\ket_{\rm cl}.\ee
In the unperturbed case, it reads (after spatial Fourier transformation)
\be 
\label{eqGR0}
G^R_0(\veck, t-t') = \theta(t-t')\frac{\sin \om_\veck (t-t')}{\om_\veck}.\ee
If one now calculates (\ref{eqS}), products of $\phi_0$ have to be 
averaged with the Boltmann weight, and for the unperturbed case this gives
\be
\label{eqS0} 
S_0(\veck, t) = \int 
d^3x\,e^{-i\veck\cdot\vecx}\bra\phi_0(x)\phi_0(0)\ket_{\rm 
cl} = T\frac{\cos \om_\veck t}{\om_\veck^2}.
\ee
As is well-known, there is a zoo of methods available to formulate 
perturbation theory in real-time quantum thermal field theory \cite{LeB}. 
When using 
the Schwinger-Keldysh contour, the quantum field $\phi$ is denoted with 
$\phi^+, \phi^-$, if it lives on resp. the upper and 
the lower branch. A version that is particulary convenient 
here, is the  'center-of-mass/relative' coordinates version, where 
the basic field variables are taken as
\[
 \left( \begin{array}{c} \phi_1\\
\phi_2
\end{array}\right) =
\left(\begin{array}{c}
(\phi_+ +\phi_-)/2
\\
\phi_+ - \phi_-
\end{array}\right),
\]
in terms of $\phi^+, \phi^-$.
The reason is that the matrix propagator in this basis compares directly 
with the 2-point functions that we found in classical perturbation theory. 
Namely, the $2\times 2$ propagator is given by
\[
{\bf G}(x-x') =
\left( \begin{array}{cc} iF(x-x') & G^R_{\rm qm}(x-x')\\
G^A_{\rm qm}(x-x') & 0
\end{array}\right),
\]
with in terms of the original field
\bean
F(x-x') &=&
\half \bra \phi(x)\phi(x')+\phi(x')\phi(x)\ket ,\\
G^R_{\rm qm}(x-x') &=& G^A_{\rm qm}(x'-x) = 
i\theta(t-t')\bra[\phi(x),\phi(x')]\ket. \eean
Indeed, in the naive classical limit, i.e. fields commute and commutators go 
to Poisson brackets, these become (\ref{eqS}) and (\ref{eqGR}).

In the unperturbed case, the relation becomes very explicit. The free $G^R_0$ 
is given by the classical expression (\ref{eqGR0}), and the free 
$F_0$ is given by 
\[ 
F_0(\veck, t) = [n(\om_\veck)+\half]\frac{\cos \om_\veck t}{\om_\veck} =
T\sum_n \frac{\cos \om_\veck t}{\om_n^2+\om_\veck^2},
\]
where $n(\om) = (e^{\beta\om}-1)^{-1}$ is the Bose 
distribution. We have written the real-time 2-point function that contains 
all the temperature dependence as a sum over Matsubara frequencies 
$\om_n=2\pi n T$, familiar from the imaginary-time formalism. Comparing 
this result with (\ref{eqS0}) indicates 
that the classical theory is an effective theory for the $n=0$ term, just 
as in DR for time-independent quantities. 

An important ingredient in the setup of perturbation theory in the 
quantum case is the KMS condition \cite{Nie}. Using the KMS condition to 
determine 
the 2-point function in temporal momentum space, ensures that the 
so-called vertical part of the contour is taken into account, or in more 
physical terms, that the system is in thermal equilibrium. It turns out 
that it is possible to derive a KMS condition also in the classical 
theory \cite{Pa}, it reads
\[ \frac{d}{dt}\bra\phi(x)\phi(x')\ket_{\rm cl} = T\bra\{\phi(x), 
\phi(x')\}\ket_{\rm cl}.\]
It can be used to express $S$ in terms of $G^R, G^A$. 
Explicitly, we find in resp. the quantum and the classical case 
the following relation in temporal momentum space ($k=(k^0,\veck)$)
\bean
F_0(k) &=& -i(n(k^0)+\half)\left(G^R_0(k)-G^A_0(k)\right),\\
S_0(k) &=& -i\frac{T}{k^0}\left(G^R_0(k)-G^A_0(k)\right).
\eean 
Using this, we can work in temporal momentum space throughout, which is 
convenient from a technical point of view.

All the found relations between the quantum and the classical expressions 
are nice, but rather academic until we start to calculate perturbative 
corrections. Hence we calculated the 2-point function to two loops and 
the 4-point function to one loop.
Both in the quantum as in the classical theory, it is straightforward 
to recognize diagrams and identify 1PI parts. It turns 
out that there is a direct correspondence with the diagrams in the 
quantum theory if one uses the $\phi_1, \phi_2$ basis. I will show this 
for the one loop and two loop setting sun retarded self-energy 
$\Sigma_R(p)$.

\subsection*{One loop self-energy}
In the quantum theory, the one loop retarded self-energy is given by (see 
Fig.\,1) 
\bea
\nonumber
 \Sigma_R^{(1)} &=& \half\lambda\int \frac{d^4k}{(2\pi)^4}\,F_0(k) \\
\label{eqs1}
&=&\frac{\lambda T^2}{24\hbar} - \frac{\lambda mT}{8\pi} +{\cal 
O}(\hbar\lambda \log (T/\hbar)),\eea
where I indicated explicitly the $\hbar$ dependence.
The leading term is the so-called thermal mass, $m_{\rm th}^2$, which is 
the only Hard Thermal Loop (HTL) contribution in scalar field theory. It 
needs to be resummed in order to have a true perturbative expansion (in 
$\lambda^{1/2}$). 

\begin{figure}
\centerline{\psfig{figure=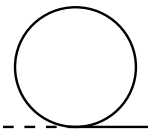,width=2.0cm}}
\caption{\narrowtext Retarded self energy: one loop diagram.
The full line represents $F_0$ resp. $S_0$ and the dashed-full line 
$G_0^R$.} \end{figure}

In the classical theory, we find (with a momentum cutoff $\Lambda$)
\be
\label{eqs2}
 \Sigma_{R, \rm cl}^{(1)} = \half\lambda\int 
\frac{d^4k}{(2\pi)^4}\,S_0(k) 
=\frac{\lambda \Lambda T}{4\pi^2} - \frac{\lambda mT}{8\pi}.\ee
Compared with the quantum expression, $F_0$ is replaced by $S_0$. 
The $\hbar^0$ term in (\ref{eqs1}) is correctly reproduced by the classical 
theory, 
and the higher order terms in $\hbar$ are absent, as expected. The 
$\hbar^{-1}$ term turns up in the classical theory as a linear 
divergence, and not as the thermal mass. However, if we use our knowledge 
from DR, we can simply cancel the divergence and put in the correct 
thermal mass by the appropriate choice of classical mass parameter, 
which is written as $m^2 = m_{\rm th}^2 -\delta m^2$. In DR, this is 
called matching. Note that since this one loop 
diagram is momentum independent, the analysis is not very complicated.

\subsection*{Two loop setting sun diagrams}
The two loop setting sun contribution to the self-energy is 
momentum-dependent and gives in the quantum theory rise to e.g. Landau 
damping, because of an imaginary part. Hence it is worthwhile to analyse 
this diagram also in the classical theory.
In the quantum theory we have two diagrams (see Fig.\,2)
\bean
\Sigma_{R}^{\rm sun}(p) = -\half \lambda^2\int
\frac{d^4k_1}{(2\pi)^4}\frac{d^4k_2}{(2\pi)^4} \Big[
F_0(k_1)F_0(k_2)G^R_0(k_3)&&\\ 
-\frac{1}{12} G_0^R(k_1)G_0^R(k_2)G_0^R(k_3)\Big], && 
\eean 
with $k_3 = p-k_1-k_2$.
Note that the second diagram does not contain (explicit) temperature 
dependence. A closer look shows that it is indeed subdominant at high 
$T$ and small $\lambda$. 
\begin{figure}
\centerline{\psfig{figure=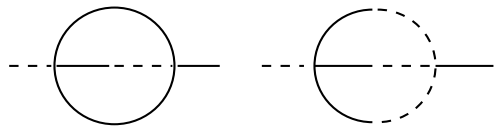,width=6.5cm}}
\caption{\narrowtext Retarded self energy: two loop setting sun diagrams. 
The second diagram is absent in the classical theory.} 
\end{figure}
It turns out that in the classical theory we only find the 
dominant contribution, i.e. the classical counterpart ($F_0\to 
S_0$) of the first diagram 
\bea
\nonumber
\Sigma_{R, \rm cl}^{\rm sun}(p) &=&
-\half \lm^2
\int\frac{d^4k_1}{(2\pi)^4}\frac{d^4k_2}{(2\pi)^4}
S_0(k_1)S_0(k_2)G^R_0(k_3)\\
\label{eqsun1}
&=&
-\frac{\lm^2T^2}{6} 
\int \frac{d^3k_1}{(2\pi)^3}\frac{d^3k_2}{(2\pi)^3}
\frac{1}{\om_{\veck_1}^2 \om_{\veck_2}^2 \om_{\veck_3}^2}\\
\label{eqsun2}
&&\;\;\;\;\;\;\;\;+\frac{p^0}{T}\int 
\frac{d\Om}{2\pi}\frac{w(\vecp,\Om)}{p^0+i\ep +\Om}, \eea
where we introduced the classical scattering integral 
\bean
 w(\vecp, \Om) &=&
\frac{\lm^2}{6}
\sum_{\{\pm\}}\int
\prod_{j=1}^3\left[ 
\frac{d^3k_j}{(2\pi)^32\om_{\veck_j}}\frac{T}{\om_{\veck_j}}\right]\\
&&\times (2\pi)^{4}
\delta(\Om \pm\om_{\veck_1}\pm\om_{\veck_2}\pm\om_{\veck_3}).
\eean
The sum is over all $+$'s and $-$'s.
The first term (\ref{eqsun1}) is independent of $p^0$, real, and 
logarithmically divergent. It is actually the result obtained in DR. 
Hence the divergence is canceled again by the appropriate choice of 
classical mass parameter, just as in the one loop diagram. The 
divergences we have encountered by now are the only ones in DR, because of 
superrenormalizability. A closer look at other diagrams in the 
classical theory makes it plausible that these divergences are also 
the only ones in the general time-dependent classical case. Hence the 
classical divergences are completely under control.
The second term (\ref{eqsun2}) is finite. It contains all the $p^0$ i.e. 
time dependence, and it has an imaginary part. 

It is instructive to compare the imaginary parts in the quantum and 
classical case. Using the notation \cite{WaHe} (for $p^0>0$)
\[ \mbox{Im}\; \Sigma_{R}^{\rm sun}(p) = -g_{1}(p) -g_{2}(p),\]
for both the quantum and classical case, 
we find, in the quantum theory
\bean
g_1(p) &=& \frac{\lambda^2}{96}(e^{p^0/T}-1)\int 
\frac{d^3k_1}{(2\pi)^3}\frac{d^3k_2}{(2\pi)^3}
\frac{n_{\veck_1}}{\om_{\veck_1}}\\
&&\times
\frac{n_{\veck_2}}{\om_{\veck_2}}
\frac{n_{\veck_3}}{\om_{\veck_3}}
2\pi \delta(p^0 - \om_{\veck_1}-\om_{\veck_2}-\om_{\veck_3})
,\\
g_2(p) &=&
\frac{\lambda^2}{32}(e^{p^0/T}-1) \int 
\frac{d^3k_1}{(2\pi)^3}\frac{d^3k_2}{(2\pi)^3}
\frac{(1+n_{\veck_1})}{\om_{\veck_1}}\\
&&\times
\frac{n_{\veck_2}}{\om_{\veck_2}}
\frac{n_{\veck_3}}{\om_{\veck_3}}
2\pi \delta(p^0 + \om_{\veck_1}-\om_{\veck_2}-\om_{\veck_3}),
\eean
and in the classical theory
\bean
g_{1, \rm cl}(p) &=& 
\frac{\lambda^2}{96}\frac{p^0}{T}
\int 
\frac{d^3k_1}{(2\pi)^3}\frac{d^3k_2}{(2\pi)^3}
\frac{T^3}{\om_{\veck_1}^2 \om_{\veck_2}^2 \om_{\veck_3}^2}\\
&&\times
2\pi \delta(p^0 - \om_{\veck_1}-\om_{\veck_2}-\om_{\veck_3})
,\\
g_{2, \rm cl}(p) &=& 
\frac{\lambda^2}{32}\frac{p^0}{T} 
\int 
\frac{d^3k_1}{(2\pi)^3}\frac{d^3k_2}{(2\pi)^3}
\frac{T^3}{\om_{\veck_1}^2 \om_{\veck_2}^2 \om_{\veck_3}^2}\\
&&\times 2\pi \delta(p^0 + \om_{\veck_1}-\om_{\veck_2}-\om_{\veck_3}),
\eean
$g_{1,(\rm cl)}(p)$ represents three body decay, and $g_{2,(\rm cl)}(p)$
Landau damping. The classical expressions are indeed the leading order 
expressions from the quantum theory. The reason is that, after 
replacement of the Bose distributions by the classical distributions, 
the resulting integrals are still finite.

To demonstrate this very explicitly, consider 
the on-shell plasmon damping rate \cite{Pa92,WaHe}. It is given by 
\[
\gamma = \frac{-\mbox{Im}\;\Sigma^{\rm sun}_{R} (\vecnul, m)}{2m} = 
\frac{g_2(\vecnul, m)}{2m}.\]
It turns out that, to leading order in $\lambda$ and $T$, 
$g_2(\vecnul, m)$ equals $g_{2,\rm cl}(\vecnul, m)$,
and we find (to leading order)
\[  \gamma_{\rm cl} = \gamma = \frac{\lambda^2 T^2}{1536\pi m}.\]
Using that $m\approx m_{\rm th}$, because of resummation in the quantum 
theory and because of  matching in the classical theory, gives for
the plasmon damping rate (again to leading order)
\[  \gamma_{\rm cl} = \gamma = 
\frac{\lambda\sqrt{\lambda\hbar}T}{128\sqrt{6}\pi}.\]
This example shows in a very explicit way that the classical theory, with 
the appropriate choice of parameters, indeed approximates the 
quantum theory. 

Let me end the discussion on scalar field theory with 
the remark that also for the 4-point function the leading order 
expressions are reproduced by the classical theory.

\section{Classical thermal gauge theory and Hard Thermal Loops}
Of course, one of the reasons the scalar theory was studied, is because 
of possible applications to the dynamics of hot gauge theories. 
We know that the DR approach has been succesfully applied to study e.g. the 
electroweak phase transition, and the presence of (non-)abelian gauge 
fields did not pose any fundamental problems. 

It turns out that, concerning the dynamics, there are such problems 
\cite{Bo,Ar}. 
The reason for this are the Hard Thermal Loops in gauge theories 
\cite{BrPi}.  
In contrast to the scalar theory, HTL's in gauge theories are momentum 
dependent, in a complicated non-analytical way. 
As a typical example, consider the longitudinal part of the gauge boson 
self-energy in e.g. scalar electrodynamics. In the HTL approximation, it 
is given by \cite{LeB}
\[ F(p^0,\vecp) = 
2m^2\left(1-\frac{{p^0}^2}{\vecp^2}\right)\left(1-\frac{p^0}{2|\vecp|}\log 
\frac{p^0+|\vecp|}{p^0-|\vecp|}\right),\]
with the prefactor
\[ m^2 = \frac{e^2}{\pi^2}\int_0^\infty dk k n(k) = \frac{e^2T^2}{6\hbar},\]
where I indicated again the $\hbar$ dependence. This term gives rise to 
e.g. the  Debye screening mass, $m_D^2 = F(0, \vecp\to 0) = 2m^2 = 
e^2T^2/3\hbar$.

In the classical theory, the Bose distribution $n(\om)$ is replaced by 
the classical one $T/\om$, which leads to a linearly divergent prefactor
\[ m_{\rm cl}^2 = \frac{e^2}{\pi^2}\int_0^\Lambda dk k \frac{T}{k} = 
\frac{e^2T\Lambda}{\pi^2},\]
where we used a momentum cutoff for simplicity. This is similar to what 
happens in the scalar theory (compare (\ref{eqs1}) and (\ref{eqs2})). 
However, now it is not possible to cancel 
the divergence with a local mass counterterm, because of the complicated 
momentum dependence. In the DR approach for static quantities, these 
problems do not arise, because for $p^0=0$, $F(0,\vecp)=2m^2$, i.e. 
momentum independent. Hence the linear divergent Debye mass in the 
effective 3-d theory can, in DR, simply be canceled by a mass counterterm.

We can conclude that HTL's in gauge theories make a straightforward use 
of classical thermal gauge theory to approximate the dynamics 
questionable. 
In the several proposals that exist to incorporate HTL effects in a 
classical-like theory, new 
local degrees of freedom are added to make up for them in one way or 
the other \cite{Bo,Mu,Mo,Ia}. The idea to add classical particles (instead 
of fields)  has been succesfully implemented numerically \cite{Mo}.

\section{Conclusions}
To summarize, we have shown how classical thermal field theory can be 
used to approximate real-time quantum field at finite temperature, for 
the $\lambda\phi^4$ case. Instead of explicitly integrating out hard 
modes to construct an effective theory for the soft modes, we showed 
that it is possible to take all the modes into account. The resulting  
Rayleigh-Jeans divergence can be dealt with in a straightforward manner, 
namely by using counterterms that are dictated by dimensional reduction. 
Furthermore, by using real-time perturbation theory, we have shown that 
the classical theory approximates the quantum one if also 
the finite part of the classical parameters are chosen according to the 
DR matching rules.

Essential is that the HTL effects in the quantum theory can be easily
incorporated in the classical theory: only the thermal mass has to be put
in. This is also the reason why the same prescription does not work for
gauge theories. Here HTL's are momentum dependent and in the classical
theory, they give rise to divergences that cannot be canceled with local
counterterms (i.e. in $3+1$ dimensions; the classical approximation does
work for gauge theories in $1+1$ dimensions, because here the classical
contribution is dominant and finite, and the HTL contribution is
subdominant \cite{TaSm98}). Adding new degrees of freedom to represent
the HTL contributions seems to be a possible way out.

\acknowledgments
This work is supported by FOM.

\end{multicols}

\begin{references}

\bibitem{AaSm}   G.~Aarts and J.~Smit, 
		 Phys.\ Lett.\ B393 (1997) 395; 
  		 Nucl.\ Phys.\ B511 (1998) 451.
\bibitem{JaNa}   W.~Buchm\"uller and A.~Jakov\'ac, 
		Phys.\ Lett.\ B407 (1997) 39; 
		B.J.~Nauta and C.G.~van Weert, 
		hep-ph/9709401. 
\bibitem{poster}  G.~Aarts and J.~Smit, {\em Non-equilibrium dynamics with 
		  fermions on a lattice in space and time}, 
		  poster presented at this Workshop, hep-ph/9809340.
\bibitem{GrMu96}  C.~Greiner and B.~M\"uller, Phys.\ Rev.\ D55 (1997) 1026.
\bibitem{GrRu88}  D.Yu.~Grigoriev and V.A.~Rubakov,
                  Nucl.\ Phys.\ B299 (1988) 67.
\bibitem{DR} 	  K.~Kajantie, M.~Laine, K.~Rummukainen and M.~Shaposhnikov,
                  Nucl.\ Phys.\ B458 (1996) 90; B466 (1996) 189,
		   and references therein.
\bibitem{LeB}     see e.g. M.~Le~Bellac, {\em Thermal Field Theory}
                  (Cambridge University Press 1996).
\bibitem{Nie}     A.~Ni\'egawa, Phys.\ Rev.\ D40 (1989) 1199;
		  T.S.~Evans and A.C.~Pearson,
                  Phys.\ Rev.\ D52 (1995) 4652.
\bibitem{Pa}      G.~Parisi, {\em Statistical Field Theory}
                  (Addison-Wesley Publishing Company,  1988).
\bibitem{WaHe}    E.~Wang and U.~Heinz, Phys.\ Rev.\ D53 (1996) 899.
\bibitem{Pa92}    R.S.~Parwani, Phys.\ Rev.\ D45 (1992) 4695.
\bibitem{Bo}	  D.~B\"odeker, L.~McLerran and A.~Smilga,
                  Phys.\ Rev.\ D52 (1995) 4675.
\bibitem{Ar} 	  P.~Arnold, D.~Son, L.G.~Yaffe, Phys.Rev.D55 (1997) 6264; 
		  P.~Arnold, Phys.\ Rev.\ D55 (1997) 7781. 
\bibitem{BrPi}    E.~Braaten and R.D.~Pisarski,
                  Nucl.\ Phys.\ B337 (1990) 569; 339 (1990) 310;
		  J.C.~Taylor and S.M.H.~Wong,   
                  Nucl.\ Phys.\ B346 (1990) 115.
\bibitem{Mu}   	  C.R.~Hu and B.~M\"uller, Phys.\ Lett.\ B409 (1997). 	
\bibitem{Mo}  	  G.D.~Moore, C.R.~Hu and B.~M\"uller, 
		  Phys.\ Rev.\ D58 (1998) 45001.
\bibitem{Ia}      E.~Iancu, hep-ph/9710543.
\bibitem{TaSm98}  W.H.~Tang and J.~Smit, hep-lat/9805001.
 


\end{references}
\end{document}